\documentclass[12pt]{iopart}

\usepackage{xspace}
\usepackage{cite}
\usepackage{graphicx,color}
\usepackage{hyperref}
\definecolor{linkcolor}{rgb}{0,0,0.6} 
\hypersetup{
	colorlinks=true, 
	pdfstartview=FitV, 
	urlcolor=linkcolor, 
	linkcolor= linkcolor, 
	citecolor=linkcolor 
}

\begin{document}

\title[Gravity detection in plants]{A new scenario for gravity detection in
	plants: the position sensor hypothesis}

\author{O. Pouliquen$^1$, Y. Forterre$^1$,  A. B\'erut$^1$, H. Chauvet$^{1,2}$,
	F. Bizet$^2$, V. Legu\'e$^2$, B. Moulia$^2$}

\address{$^1$Aix Marseille Univ, CNRS, IUSTI, Marseille, France.\\
	$^2$Integrative Physics and Physiology of Trees (PIAF), INRA, Univ.
	Clermont-Auvergne, 63000 Clermont-Ferrand, France.}
\ead{olivier.pouliquen@univ-amu.fr}
\vspace{10pt}
\begin{indented}
	\item[]December 2016
\end{indented}

\begin{abstract}
The detection of gravity plays a fundamental role during the growth and evolution of plants. Although progress has been made in our understanding of the molecular, cellular and physical mechanisms involved in the gravity detection, a coherent scenario consistent with all the observations is still lacking. In this perspective paper we discuss recent experiments showing that the response to inclination of shoots is independent of the gravity intensity, meaning that the gravity sensor detects an inclination and not a force. This result questions some of the commonly accepted hypotheses and leads to propose a new ``position sensor hypothesis''. The implications of this new scenario are discussed in the light of different observations available in the literature.
\end{abstract}

%
%
%
%
\ioptwocol

\section{Introduction}

Gravity perception by plants plays a key role in their development and
acclimation to their environment, from the direction of seed germination to the
control of the posture of adult plants. This can be demonstrated by the ability
for the shoot to recover a vertical posture independently from light clues when 
inclined to different angles. This ability is broadly observed among the
plant world, from small wheat coleoptiles to trees \cite{bastien2013} (Fig~\ref{fig:intro}a and b).  Roots also are sensitive to
gravity and can adjust gravitropically their direction to grow deeper in the
soil (Fig.~\ref{fig:intro}c). This phenomenon has been named gravitropism (from  \textit{gravi}, gravity and the Greek \textit{tropein}, to turn). The observation and the study of gravitropism go back to the 19th century
with the pioneering work of Julius von Sachs (1868) and of the Darwins (1880)
(reviewed in \cite{firn1980, whippo2009, moulia2009}). More than a century later, our understanding has
significantly improved, but many questions remain open
\cite{morita2010,moulia2009,gilroy2008}.

Among the key issues is the sensing mechanism.
A candidate model for the sensing mechanism should meet two requirements: 1)
It should explain how cells sense their change in orientation at the cellular level. It should explain in
particular what organelles are involved, what variable is sensed
and leads to the primary physiological reaction, what molecular players are
involved, and what is the timing of the different phases. 2) It should be
consistent with the characteristics of the macroscopic response of the plant, such as the characteristic times to start
the response and to converge to the vertical, the  influence of the angle of
tilt and of the duration of the tilting stimulus  \cite{perbal1976, perbal2002}. The research tactic to achieve the
identification of the gravisensing mechanisms has thus involved studies at both
the cellular level and the macroscopic level of the organ. Currently the prevailing scenario is the following.

\subsection{The prevailing scenario}
\begin{figure*}
	\centering
	\includegraphics[width=0.90\textwidth]{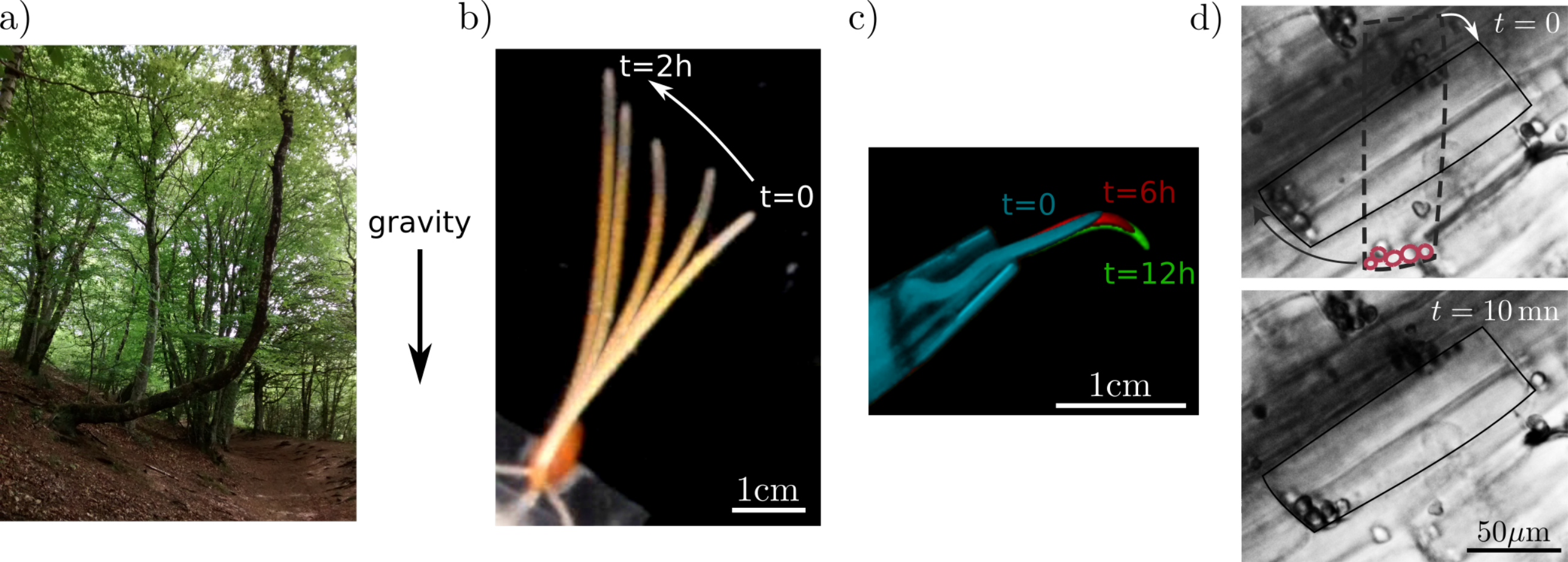}
	\caption{Illustration of gravitropism: a) on a tree, b) on a wheat coleoptile, c) on a lentil root; d)
		pictures of a statocyte (delineated in black) in wheat coleoptile showing the position of the
		statoliths just after inclinaison and $10$ mn later. Their previous positions are delineated with dotted red lines.}
	\label{fig:intro}
\end{figure*}

The perception of gravity
starts in specific cells that act as statocytes (from \textit{stato}, static position and
\textit{cyte}, cell). Indeed the suppression of these specific cells strongly inhibits
gravitropism \cite{fukaki1998, blancaflor1998}. Statocytes are located in different organs.
In coleoptiles they are found in a thin layer near
conducting tissues; in young stems, within a thin layer near the endodermis; in roots, within few columns of cells located in the
central root cap, called the columella; and statocytes have recently been localized in the secondary phloem of mature woody stems~\cite{gerttula2015}. These cells contain specialized
organelles called statoliths. Being denser than the surrounding intra-cellular
fluid, the statoliths move in the direction of the gravity Fig.~\ref{fig:intro}d and exert a force, presumably on the plasma membrane, that provides the information about the direction of gravity (the
``statolith hypothesis") \cite{morita2010}.  When the orientation of the organ with respect to the gravity vector changes, the statoliths change position, exert force on a new part of the cell, which in turn induces the relocalisation of membrane transporters called PIN proteins. The PIN proteins then redirect the flow of auxin, the major plant hormone, leading to a differential growth between the two faces of the organ and ultimately to the organ bending back to the desired orientation with respect to gravity.  (Chodlony-Went hypothesis) \cite{firn1980}.

Although some consensus has been reached about the main scenario described previously, many questions remain open, from the early steps of gravity sensing  to the auxin pathway leading to differential growth. In this perspective paper, we focus on the  gravity detection mechanism, reviewing results obtained both at the cellular and macroscopic plant scale (for a focus on the more dowstream events involving auxin transport and growth response, see \cite{vanneste2009}). 

\subsection{Open questions and hypotheses at the cellular and macroscopic scales}
At the cellular scale, the exact nature and roles of the statoliths, and whether or not
they are necessary for a graviperception is still a matter of debate. It is
generally accepted that the amyloplasts (i.e.\ organelles filled with starch
grains) are the main statoliths in the statocytes (the ``starch-statolith
hypothesis" \cite{nemec1900, haberlandt1900,morita2010}). Indeed in experiments using mutants deprived of starch
and displaying little if any cellular sedimentation of amyloplasts, the macroscopic
response to a gravi-stimulation is dramatically diminished compared to the wild
type \cite{pickard1966,kiss1989,weise1999,fitzelle2001}. However, a  response still exists suggesting that
amyloplasts may not be the sole statoliths or that they act in  gravity
detection as ``enhancers" without being necessary. On the
other hand, experiments carried out by artificially moving the statoliths using
strong magnetic field gradients \cite{kuznetsov1997, hasenstein1999}
unambiguously show that the plant bends in the direction of the displacement.
Another study implicated statoliths by using mutants having
a rigid vacuole \cite{toyota2013}. The rigidity of the vacuole prevented the
statoliths from moving and as a result, the response of the mutant dramatically
decreased. In conclusion, there is no doubt that amyloplasts convey information about the gravity field, but other actors may also play this role. 
Mitochondria and Golgi apparatus have been
shown to sediment (although much more slowly than statoliths) in oat coleoptiles and have been proposed as
secondary statoliths  \cite{shen1970},  although this point is still disputed. Other authors called for an alternative
mechanism resulting from the sensing of the pressure due to the weight of the
cytoplasm acting on the membrane or cell wall  \cite{weise1999}  (the ``protoplast pressure hypothesis''). The two
pathways (the statolith-enhanced pathway and the protoplast pressure pathway)
may even co-exist~\cite{barlow1995}.

The nature of the signal detected by the statocytes is another fundamental
question which remains largely open. And, here also, different hypotheses have been
developed. Some authors argue that the statocytes are sensitive to the pressure
exerted by the statoliths on the membrane or on the endoplasmic reticulum
\cite{darwin1903, leitz2009}. This could be achieved as the statoliths promote the opening
of mechano-sensitive ion channels, either directly or through interaction with
the actin cytoskeleton \cite{hasenstein2009, yoder2001, perbal2003}. A second hypothesis relies on the role of
protein complexes (called TOC) located in the envelope membrane of plastids which could contribute by functioning as a gravity signal transducer \cite{strohm2014}.  In this case, the proximity between the TOC and interactors in the plasmic membrane is the only requirement ; no pressure is needed. But only the first hypothesis has been fully tested in the framework of macroscopic reactions (sine law, $g\times t$ transients... as detailed later) \cite{kiss1997}. 

During the sedimentation of statoliths, certain physiological reactions have been documented in hypocotyls and in roots:  a change in apoplast pH \cite{fasano2001}, an increase in
intracellular reactive oxygen species concentration \cite{joo2001} and changes in cytosolic calcium concentration \cite{toyota2008}. Even if the involvement of these actors in gravity signaling pathway is sustained, the relationship
between them is also not completely deciphered \cite{kolesnikov2015}. At the end of the signaling
pathway, the differential cell elongation leading to bending is triggered by
redirected auxin flux to the lower side of the plant organ \cite{band2012}.
Numerous studies have showed that the directionality of auxin flow within tissues is
determined by a polar cellular localization of auxin export carriers, in
particular PIN-FORMED (PIN) proteins \cite{petravsek2006, wisniewska2006}. After an inclination of roots or hypocotyls, the
localization of the PIN proteins, mainly PIN3, becomes repolarized,
\cite{friml2002, kleine2010, rakusova2011}, leading to the redirected
auxin flux. However the
mechanisms describing how the sedimentation of statoliths triggers a relocalisation of PIN
have not yet been elucidated.

Surprisingly, questions are also open at the organ scale. Even though a precise
knowledge of the response at the organ level might help discriminate between proposed explanations, only few studies have quantitatively looked at the bending
kinematics and at its sensitivity. They fall into three very
different approaches. In the first type, the focus is on the
response of the organ to different inclinations from the vertical. The response
varies linearly with the sine of the inclination angle, a relation known as the
``sine law" \cite{dumais2013}. However, despite its naming as a ``law" and its
popularity in textbooks, few reliable measurements and
assessments of the ``sine law"  can be found in the literature, and concern a very
limited number of organs and species (\cite{iino1996,perbal1976,galland2002}, see
\cite{moulia2009} for a review). In the second type, the focus is
on the response to transient exposure to gravity. In this case the gravity is
``switched on and off" either using microgravity experiments in which transient
gravity is mimicked through centrifugation, or using clinostat experiments on
Earth  in which a transient compensation of the gravity sensing is provided by
changing the plant inclination continuously before significant statolith
downward motion can occur \cite{gilroy2008,brown1995}. 
Measurements suggest that the response to transient gravity is proportional to the dose,
namely the product of the gravity intensity with the time of exposure (the
``reciprocity rule"). Finally, the third type relies on varying the mass of
the amyloplasts in the statocytes either through drastic hormonal treatments
\cite{pickard1966}, or through a range of starch-less and starch-excess
mutants \cite{kiss1997} and studying their response to tilting (on
Earth). It was found that the macroscopic responses increases with the mass of
the statoliths. 
 
\subsection{The standard synthesis: the ``gravity-force sensing hypothesis" and the ``starch-statolith weight model"... and its call into question}

Combining such a wide range of phenomena and hypotheses for gravity sensing at the cellular and macroscopic scales into a single framework has remained a challenge. However a standard synthesis has been widely accepted (more or less implicitly) to explain the cellular and macroscopic responses related to processes upstream of the development of a gradient in auxin concentration. It is based on a  ``gravity-force sensing hypothesis" \cite{barlow1995}. Its most accepted version is the ``starch-statolith weight model", which stipulates that the statoliths sediment after tilting, and that the detected signal is the force exerted by the sedimented or sedimenting statoliths on the lateral side of the cell, or on the side of internal structures like the cytoskeleton, or the endoplasmic reticulum. In this hypothesis, gravisensing should then depend on the mass of the statoliths and on the gravity intensity. 

In this perspective paper, we discuss recent results we have obtained on shoot
gravitropism, showing that the response of a shoot is insensitive to the gravity
intensity and, hence, insensitive to the weight of the statoliths or protoplast, but solely
depends on the inclination between the shoot and the gravity vector
\cite{chauvet2016}. The gravity sensor in plants thus works as an inclination
sensor and not as a force sensor. This finding dismisses the ``gravity force
sensing" paradigm, requiring an alternative ``position sensor model" to unify 
the cellular and macroscopic results.  

At the cellular level, we point out
that the collective motion of the statoliths during tilting experiments on Earth
is actually not the sedimentation of a suspension, but the dynamics of a
grain-pile.  We discuss the strong implications of these two results and how
they help us to revisit the gravisensing pathway. We then propose an
alternative theory,  the ``statolith position hypothesis". Our hypothesis is
that the relevant parameter sensed by the statocytes is the position of the
statolith pile. In a vertical organ, the statoliths settle down  and form a
pile at the bottom of each cell. When inclined, they avalanche on the lower
side of the cell, and their new position provides the information about the
direction of gravity. We will show that this ``position sensor model" can provide
an explanation for the sine law and also gives an interpretations for the
response to transient stimuli, the reciprocity rule, and may also account for the phenotype of the starch mutants. We finally speculate
about a possible mechanism to connect this statolith-pile position with the
distribution of the auxin transporters (PINs), and the onset of a lateral gradient in auxin
concentration.

\section{Response of a shoot to permanent stimuli: influence of  gravity intensity}

A shoot initially inclined bends up actively and goes back to its original orientation with respect to gravity (often but not always vertical).
Having a proper quantitative knowledge of the response is crucial to phenotype
mutants and to compare this phenotype with microscopic modeling of 
statolith motion and sensing. However, relatively few studies address the issue
and they are often difficult to compare. One difficulty comes from the choice of
a quantitative measurement of ``the gravitropic response"\cite{moulia2009}. Some authors have
chosen to measure the angle of the tip of the stem after a given time
\cite{iino1996,tanimoto2008}, while others measure the differential growth
between the top and the bottom face of the inclined stem \cite{myers1995}, or use
the temporal variation of the angle \cite{bastien2013}. 
Another difficulty to properly measure the gravitropic response comes from the
fact that the movement is not controlled solely by gravity perception. A
cross-talk with a second concurrent sensing comes into play: the sensing by the
organ of its own local curvature independently of gravity, called proprioception
\cite{hamant2016}.  A minimal unifying model for the combined
control through gravisensing and proprioception has been developed and validated
on  many species \cite{bastien2013,bastien2014}. This approach opened new
perspectives and suggested that the response to inclination should be revisited
and that care should  be taken when analyzing the macroscopic response. Measurements
should be done when gravisensing is dominant compared to proprioception, which
means when the curvature is negligible, i.e. at the beginning of the bending.  It also
means that measure should be done on organs sufficiently stiff so that they do not
bend under their weight when inclined.  

To go beyond the specificity of each species and
find general tendencies,it is useful to work with a defined dimensionless number. With these precautions in mind, it is possible to define a relevant
dimensionless measure for the gravitropic response of shoots, as proposed by
Bastien et al. \cite{bastien2013,bastien2014}. The dimensionless gravitropic response number introduced by
Bastien et al. relies on the comparison of the bending velocity to the growth
velocity. In the following, we explain how this definition can be derived from kinematic arguments,  before
presenting the response of shoots to permanent stimulus at various
inclinations and gravity intensities and discussing the so called ``sine law".  

\subsection{A valid measure of the gravitropic response}

To define a valid macroscopic quantification of response of an organ (shoot or root)  to a change in gravity condition,
we treat the growth mechanism at the origin of
bending. Let us consider an initially straight stem, for which there is no
proprioception. Once the shoot is inclined, the change in gravity direction
induces an asymmetric flux of auxin and differential growth. A relevant
response would then correspond to the measure of the relative asymmetry in
auxin concentration, namely  the difference in auxin between the top and the
bottom sides compare to average concentration of auxin. This dimensionless response $\tilde
\Delta$  can then be written as:

\begin{equation}
\label{eq:defdelta}
\tilde \Delta=  \frac{\phi_{\mathrm{low}}-\phi_{\mathrm{up}}}{\phi_{\mathrm{low}}+\phi_{\mathrm{up}}}
\end{equation}

\noindent where $\phi_{\mathrm{low}}$ is the auxin
concentration in the bottom half, and  $\phi_{\mathrm{up}}$ the auxin concentration in the top half of the stem.

\begin{figure}
	\centering
	\includegraphics[width=1\linewidth]{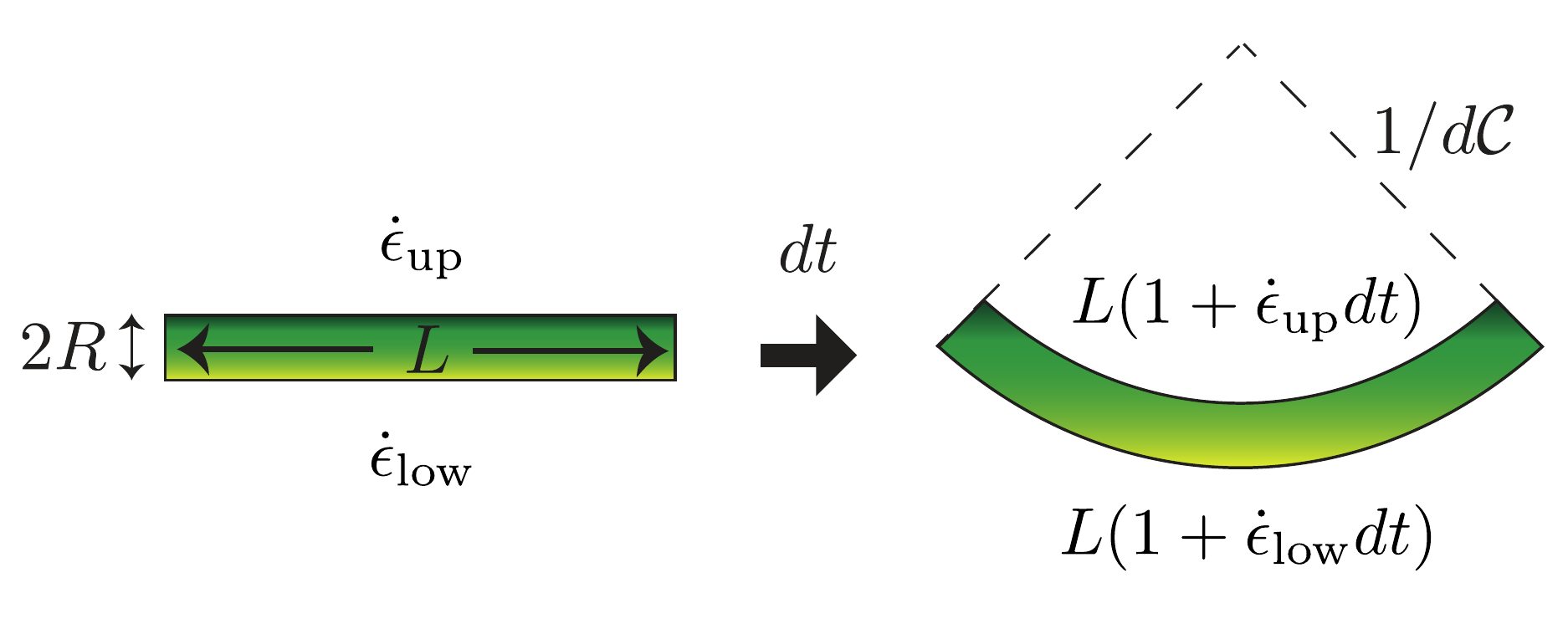}
	\caption{Bending resulting from the relative growth between the two sides of a beam. }
	\label{fig:bendig}
\end{figure}

It is known that the growth rate in shoots is an increasing function of the auxin concentration until an optimum concentration is reached \cite{galston1949,hopkins2008}. Therefore,  
in a finite range of auxin concentration, it is possible to linearize the law and to assume that the elongation
rates $\dot \epsilon_{\mathrm{low}}$ and $\dot \epsilon_{\mathrm{up}}$  (i.e.\ the relative change in length per unit of time
on each side) are proportional to the concentration of auxin:
\begin{equation}
\label{eq:epsilon}
\dot \epsilon_{\mathrm{low}}=k \phi_{\mathrm{low}} \quad \mathrm{and}  \quad \dot \epsilon_{\mathrm{up}}=k \phi_{\mathrm{up}} .
\end{equation}
The
proportionality factor $k$ may depend on genetic factors or environmental
factors such as temperature.
The difference in elongation rate between the two sides induces a bending as
shown in Fig.~\ref{fig:bendig}. An initially straight piece of shoot of length
$L$ and diameter $2R$ ends up after a time $dt$ in a curved cylinder
characterized by a length at the bottom equal to $L(1+\dot
\epsilon_\mathrm{low}dt) $, and a lengh at the top equal to $L(1+\dot \epsilon_\mathrm{up}dt)$. One can then easily show
that the difference in lengths induces a curvature $d\cal C$  given by  
\begin{equation}
\label{eq:curvature}
2 R d{\cal C}=dt(\dot \epsilon_\mathrm{low}-\dot \epsilon_\mathrm{up}).
\end{equation}
Using equations \ref{eq:defdelta}, \ref{eq:epsilon} and \ref{eq:curvature}, the
response $\tilde \Delta$ can then be expressed as a function of the mean
elongation rate $\dot \epsilon_\mathrm{mean}=(\dot \epsilon_\mathrm{low}+\dot \epsilon_\mathrm{up})/2$:
\begin{equation}
\label{eq:properresponse}
\tilde \Delta= \frac{R d {\cal C}/dt}{\dot \epsilon_\mathrm{mean}}. 
\end{equation}

The measurement of $\tilde \Delta$ requires the measurement of the rate of
change in curvature along the stem $dC/dt$ and of the
relative elemental growth rate $\dot \epsilon_\mathrm{mean}$ at each position along the reacting
organ \cite{silk1984,moulia2009}. Curvature and elongation changes can be
measured through image analysis \cite{bastien2013,bastien2014} and recently an
automatic and user-friendly tool has been released to do so \cite{bastien2016}.
However, as it involves double spatial derivative and a time derivative, these
measurements are extremely sensitive to noise, and require
cumbersome experiments. This limit can be overcome by considering the averaged
curvature and averaged growth rate in the growth zone of length $\ell$. The
curvature can be approximated by ${\cal C}\approx
(\theta_{tip}-\theta_{base})/\ell$, where $\theta_{tip}$ and $\theta_{base} $ are
the angles at the tip and at the base of the stem, and the growth rate can be
approximated by $\dot \epsilon_\mathrm{mean}\approx (dL/dt)/\ell$ where $L$ is the length of the stem.
The gravitropic response is then given by:
\begin{equation}
\label{eqn:graviresponseadim}
\tilde \Delta=R\frac{d\theta_{tip}/dt}{dL/dt}
\end{equation}
because $\theta_{base}$ is independent of the time.\\
Therefore, the relevant gravitropic response is made dimensionless by
comparing the speed of
the bending to the speed of growth. Measuring the velocity at which the shoot
comes back to vertical, i.e.\ $d \theta_{tip}/dt$, as done in several studies
\cite{iino1996}, is thus a good estimate of the gravitropic response, but only if
the growth rate remains constant. A shoot growing two times faster because of
environmental or genetic changes will come back to the vertical two times
faster, but it does not mean that the plant is twice more sensitive to gravity,
as noticed by some authors \cite{pickard1966}. This trivial effect linked to the
growth velocity is properly taken into account by using the dimensionless
response $\tilde \Delta$.  The
relevance of this formulation has been tested by Chauvet et al.
\cite{chauvet2016} where it was shown that changing the temperature of the
growth chamber modifies the growth velocity and the bending velocity but not the
ratio of the two, meaning that the gravitropic response is the same. A last
important remark is that the above analysis only holds at the first instants of
the bending, when the curvature is sufficiently small so that the proprioception
can be neglected. 

\subsection{Response to a permanent stimulus: the sine law.}

\begin{figure*}
	\centering
	\includegraphics[width=.9\textwidth]{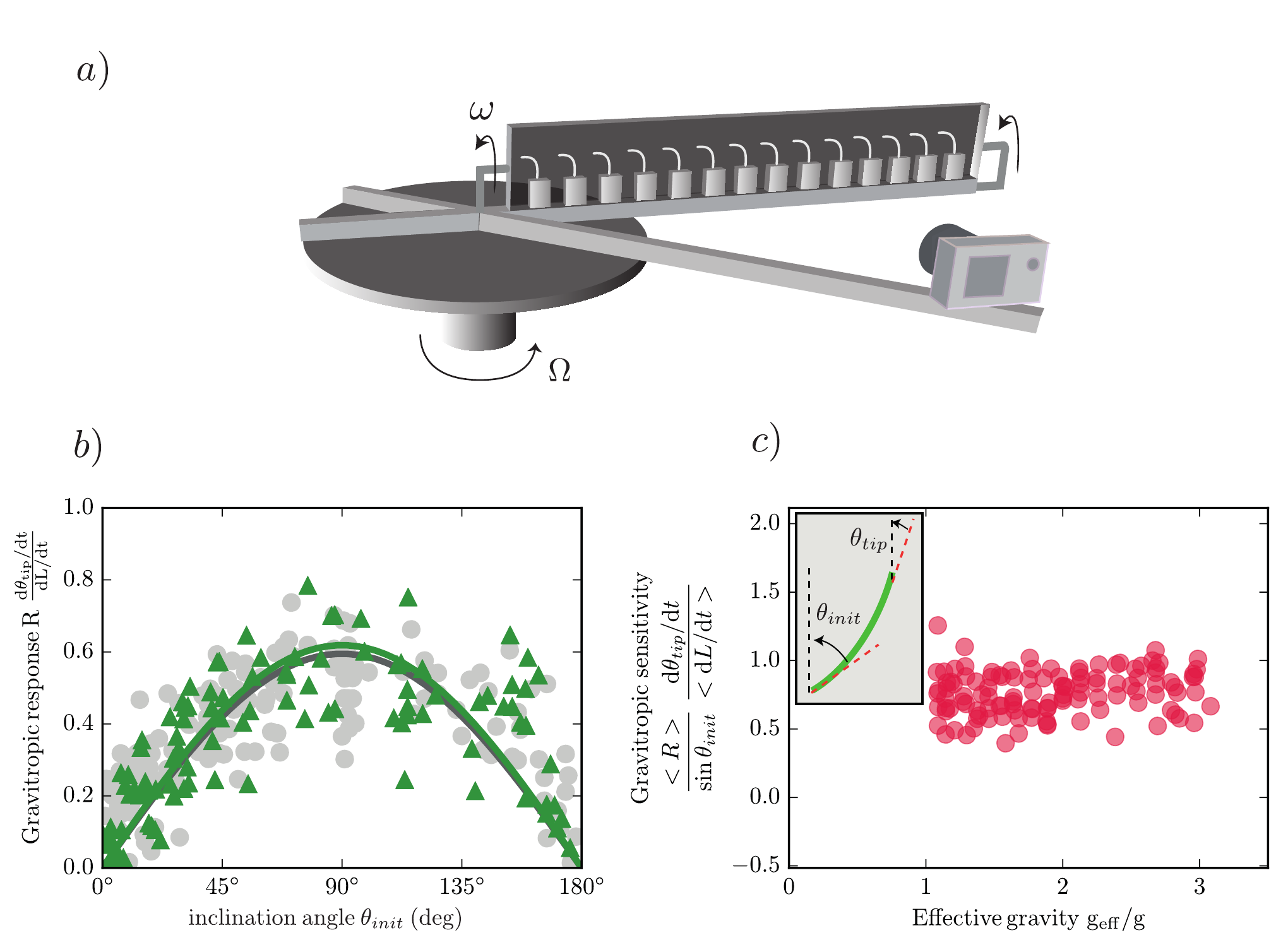}
	\caption{ a) Sketch of the experimental setup: a clinostat rotating at a slow
		rotation velocity ($\omega$) is fixed on a table rotating at a fast rotation speed ($\Omega$). The slow rotation disrupts the response to Earth's gravity and the fast rotation creates an effective gravity $g_{eff}$ which increased with the distance from the center of the rotating table and its rotation speed. Plants are grown in individual small boxes loaded into the clinostat and their kinematics induced by the local $g_{eff}$ is recorded using a camera synchronised with the clinostat rotation ($\omega$).The sine law for wheat coleoptiles; b) gravitropic response as a
		function of the inclination angle $\theta$ for gravity intensities
		equal to 1 (gray symbols) and 2.5 (green symbols) the earth gravity (data from \cite{chauvet2016}), c)
		gravitropic response normalized by the sine of the inclination angle as a
		function of the effective gravity (data from \cite{chauvet2016}).}
	\label{fig:sin}
\end{figure*}


Having defined a valid measure of the gravitropic response, one can now analyse
how the response depends on the stimulus applied to the shoot. The simplest
experiment consists in studying how a shoot goes back to the vertical when
initially inclined at an angle $\theta_{init}$ from the vertical. In this
configuration and using the dimensionless response $\tilde \Delta$,  the
``sine law" has been observed on a set of species sampling the major clades of angiosperms (flowering plants) namely rosids, asterids and commelinids \cite{chauvet2016}. The law stipulates that
the response of the shoot inclined at an angle from vertical varies linearly
with the sine of the inclination angle \cite{iino1996}. 

The results are presented in Fig.~\ref{fig:sin}b for wheat coleoptiles (grey
symbols). When increasing the initial inclination angle, the gravitropic
response increases roughly linearly, reaches a maximum around 90 degrees when
the plant is put in a horizontal position, and decreases when the shoot is
further inclined upside down. Data are well fitted by a sine law. The novelty
of the study was to investigate the influence of the gravity intensity $g$ on
the response, independently of the inclination angle $\theta_{init}$. This was achieved by developing a growth chamber on a rotating table (Fig.~\ref{fig:sin}a) able to induce
an additional centrifuge acceleration and to mimic hyper gravity conditions
under an effective gravity $g_{\rm eff}$. Data in green in Fig.~\ref{fig:sin}b
have been obtained for an effective gravity $g_{\rm eff}=2.5 g_{\rm earth}$ and
follow quantitatively the same ``sine law" as for $g_{\rm eff}=g_{\rm earth}$.
This independence holds over a whole range of gravity intensity as shown in
Fig.~\ref{fig:sin}c where the response normalized by $\sin \theta_{init}$ is
plotted versus $g_{\rm eff}$. 

The independence of the gravitropic response with
gravity intensity has also been observed for different organs from species
broadly representative of land angiosperms and in hypo-gravity condition down to
$0.1 g_{\rm earth}$ using a specific clinostat mounted on a rotating table (Fig.~\ref{fig:sin}a). Counterintuitively, this study thus concludes that
gravisensing in plants appears to be independent of the amplitude of gravity. It
implies that gravity sensing in plants works as a clinometer, sensing
inclination angles, and not as an accelerometer or a force sensor. In this
sense, the gravity sensor in plants contrasts with the inner ear system of
vertebrates, which is based on the deflection of ciliar cells induced by the
force exerted by otoliths (small grains attached to the cells). Whereas
vertebrates do not detect the difference between transient accelerations and 
inclinations, plants do, which may be a good strategy to be less sensitive to
vibrations induced for example by wind. In the next section we discuss how this
observation has implication for the understanding of the gravity
sensor at the cellular level.

\section{The Position Sensor Hypothesis}

The observations reported in \cite{chauvet2016}, and discussed
in the previous section, strongly constrains the various hypotheses of gravity detection at the cellular level discussed in the introduction, as they show that the response is insensitive to the gravity
intensity and only varies with the inclination angle. 
More precisely, a gravitropic response independent of $g_{eff}$ over a wide range dismisses the
scenarios based on the measure of the statoliths weight, such as the
assumption  that  the signal is triggered by the pressure exerted by the
statolith on the endoplasmic reticulum or on the membrane, or the assumption
that the weight of the statoliths induces a deformation of the actin network and
activates the signal. If the weight of the statoliths was the measured signal,
the gravitropic response should depend on the gravity intensity.  By the same
token, it discards the ``protoplast pressure hypothesis" which has been proposed
as an alternative candidate mechanism to explain the behavior of starchless
mutants. 

The result presented in section 2 thus strongly suggests that the relevant
stimulus is the position of the statoliths. When the shoot is inclined, the
final position of the grains within the cell is independent of the gravity
intensity but only depends on the inclination. As long as the gravity is not
zero, the statoliths will move to the lower corner of the cell (Fig.~\ref{fig:intro}d). The averaged position of the statolith pile is thus a good
candidate for the relevant stimulus detected by the statocytes. This idea is also supported by other experiments, showing that no gravitropic response is observed if the displacement of statoliths is impeded by a rigid vacuole \cite{toyota2013}, or showing in roots that the asymmetry in auxin fluxes is correlated to the motion of the statoliths \cite{band2012}. Therefore, we propose
a  ``position sensor hypothesis" stipulating that the signal controlling the
gravisensing is the position of the statolith assembly within the cells. 

This hypothesis raises several questions and needs to be confronted with our current
knowledge of the response of plants to different stimuli.  In the following we
discuss the implications of the ``Position Sensor Hypothesis"  for
our understanding of the gravisensing chain. A first question concerns the
angular sensitivity of plants, which are able to respond to small inclination
angles ($\leq 10^\circ$). How a position sensor can be so sensitive is a non trivial question.
A second question concerns the origin of the ``sine law", and how it can be
explained within this hypothesis. The third point concerns the response to
transient stimuli and how the position sensor hypothesis may reconcile the ``sine
law" and the ``reciprocity rule". The last question concerns the response of
starch-less or starch-excess mutants, and whether or not the observation of their response
might be compatible with the ``Position Sensor Hypothesis".

\section{Implications of the Position Sensor Hypothesis} 

\subsection{Sensitivity to small inclinations: a liquid behavior of the	statoliths?}

Assuming that the position of the statoliths is the relevant parameter implies
that the statoliths have to move and change position to induce a signal. From a
physical point of view this is not as trivial as it sounds. Statoliths are not
isolated elements free to move in a clear fluid. First, they are embedded in a
complex and highly heterogeneous medium (the cytoskeleton and the vacuole) whose
mechanical properties are not well understood.  Second, when the plant is
upright, they form a dense assembly of particles at the bottom of the cell in
which the motion of one grain can be strongly affected by the surrounding
grains. Their motion is thus collective and may be compared to a submarine granular avalanche \cite{sack1985,sack1986,leitz2009}. Everyday experience tells us that a packing of grains resting at the
bottom of a container full of liquid will not move when the container is
inclined, unless the inclination becomes higher than a critical angle, typically
around 25 degrees for spherical grains. In the
physics of granular media, this is called the pile angle, and reflects the difficulty for the grains to flow
due to the friction between them and to the geometrical entanglement in the packing
\cite{andreotti2013}. If statoliths were behaving like simple passive grains
(like  sand grains), they would not move when the plant is inclined at an
angle less than 25 degrees and the ``position sensor hypothesis'' would then predict
no gravitropic response. A plant inclined at small angle would grow inclined
without ever reaching the vertical, meaning that an angle threshold would exist
in the detection. This is not what is observed, as evidenced by the sine law in
Fig.~\ref{fig:sin}b showing that even at small angles a response exists and
that the plant indeed comes back to the vertical. 

However, statoliths are not
sand grains and may not behave like passive grains. When
observing the dynamics of the statoliths in the statocytes, statoliths seem to
follow erratic trajectories and look as though they were constantly agitated \cite{sack1986,leitz2009,nakamura2011,zheng2015}. This
constant agitation arguably helps the grains to rearrange and to move with respect to their neighbors, even at small inclinations. The pile of agitated statoliths could then
flow like a liquid to the lower corner of the cell
and may not behave like a classical granular medium. The remarkable sensitivity
of plants at low inclination angles could thus be explained within the ``Position
Sensor Hypothesis",  thanks to the  agitation of the statoliths.

In this scenario, the origin of the agitation becomes an important question that
should be studied further. Some studies
\cite{sack1986,saito2005,leitz2009,nakamura2011} suggest  that statoliths experience two kinds of fluctuating motions:
large saltation motion together with small vibration-like dynamics. The fluctuations have been analyzed recently using tools from statistical physics of thermal colloidal systems, assuming that the agitation was of Brownian origin \cite{zheng2015}. However, the
size of the statoliths being relatively large (between 3 to 8 $\mu$m), it is not obvious that thermal fluctuations are sufficient to explain the erratic motion observed in the statocytes.  

The
cell activity, and more precisely the dynamics of the actin cytoskeleton
\cite{guo2014} is a good candidate as a source of agitation for the
statoliths. Indeed, experiments using actin inhibitory compounds show that the
cytoskeleton actually plays a role although the reports are contradictory
\cite{Blancaflor2013}. The response of stems
\cite{yamamoto2002,hou2004,nakamura2011} or roots \cite{mancuso2006} when
inclined to the horizontal seems faster and stronger using drugs that depolymerize
the actin network, whereas drugs that stabilise polymerization or prevent
actin depolymerisation decrease the response \cite{mancuso2006}. This
observation could be rationalized within the ``Position Sensor Hypothesis". At high
inclination, the cytoskeleton and its permanent activity actually
hinders the avalanche of the grains and may slow down their displacement and
resuspend the statolliths \cite{friedman2003,volkmann1999}, playing the role of
an inhibitor for gravity perception. At small inclination by contrast, the activity
facilitates the motion of the grains that would be jammed otherwise. The
scenario that the gravity sensor is based on the avalanches of an active
granular medium needs to be confirmed by more detailed studies of the motion of
the statoliths, of their agitation and their interaction with the cytoskeleton.

\subsection{The sine law}

\begin{figure*}
	\centering
	\includegraphics[width=0.80\textwidth]{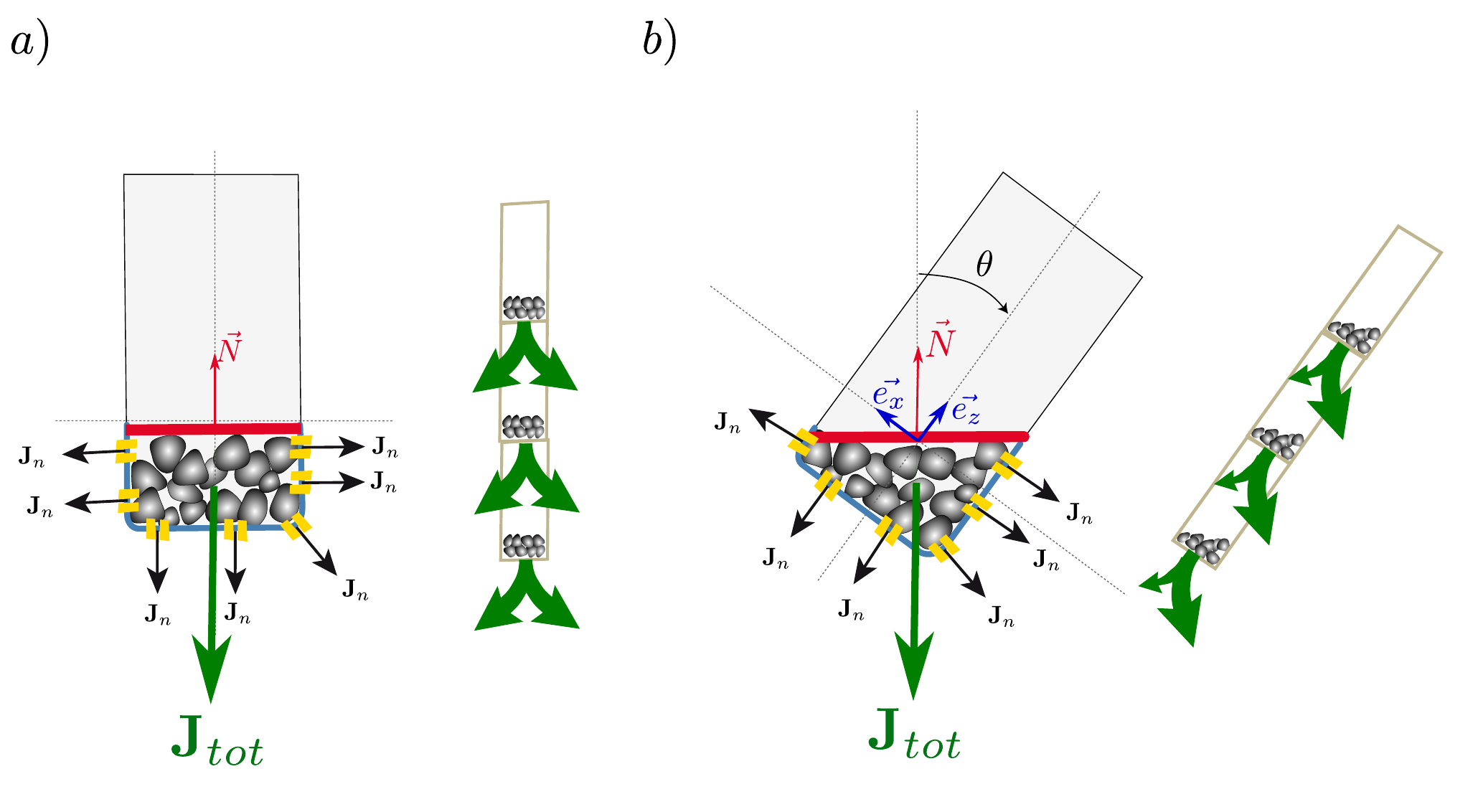}
	\caption{Sketch illustrating the scenario proposed for the ``position sensor hypothesis'':
		the proximity or the contact between statoliths and the membrane induces local
		auxin fluxes (black arrows). The resultant total flux (green arrows) is thus
		aligned with the cell axis when the cell is vertical (a), but present an
		assymetry when the cell is inclined (b). The red line represents the free surface
		of the statolith pile.}
	\label{fig:auxin}
\end{figure*}

One of the most robust results on the gravitropic response is the ``sine law"
presented in section 2. The text-book explanation for the variation of the
response as a function of the inclination angle relies on a force argument: the
statoliths exert a lateral force on the side wall of the cell, which is
proportional to $M_b g \sin \theta$, where $M_b$ is the buoyancy-corrected mass
of the statoliths. However, this explanation does not fit with the recent
observation that the gravitropic response is independent of the gravity
intensity. Thus, it is legitimate to wonder how the ``sine law" may be explained within a scenario where the relevant stimulus is the
position of the statoliths.  

A plausible mechanism is described in Fig.~\ref{fig:auxin}. Let us speculate that the proximity or the contact of the grains
with the membrane induces a flux of auxin \cite{kleine2010,dharmasiri2006}. One
can think of a mechanism involving a key-lock system or steric
interactions which may locally perturb most membrane
trafficking. When the plant is in a vertical position, the statoliths have
sedimented at the bottom of the cell, and the local auxin fluxes at each contact
have a symmetric distribution between the two lateral sides of the cell as drawn
in Fig. \ref{fig:auxin}a. The resulting  total flux is thus vertical, aligned
with the longitudinal axis of the cell, which induces no differential growth
between the two sides.  When the cell is inclined at an angle $\theta$, the
statoliths flow in the lower corner and form a pile with a free surface which is
horizontal thanks to the liquid-like behavior previously discussed. In this
configuration, there are more contacts between statoliths and the membrane on
the lower side of the cell than on the top side, and thus more local fluxes of
auxin. If one assumes that the elementary flux per unit of contact area is
uniform, the integral over the total contact area gives  by geometrical argument
a total flux which is aligned with the normal $\vec N$ to the free surface. The
free surface being horizontal, the total flux is then again aligned with the
vertical. The total flux $\bf{J}_{tot}$ is thus no longer aligned with the axis
of the cell but makes an angle $\theta$ with the cell longitudinal axis. The lateral
component of the flux along the ${\bf e}_x$ direction is then proportional to
the sine of the inclination angle $\theta$. This lateral flux could then give
rise to a differential growth and to the gravitropic response at the plant
scale. In this cartoon where the proximity of the statoliths to the membrane is
the stimulus, the ``sine law" results from the geometrical asymmetry of
the position of the statoliths in the cell since the lateral flux evolves proportionally to $\sin \left( \theta \right)$.  A way to further investigate the
relevance of this scenario would be to look for a ``sine law" in the transduction
signal itself, for example in the auxin distribution or in the  PIN
relocalisation \cite{friml2002}.  In any case, the position sensor hypothesis
provides a new interpretation of the ``sine law", that is consistent with the
insensitivity of gravisensing to the intensity of the gravity vector.

\subsection{The reciprocity rule}

We now discuss the second type of experiment focused on the response of
plants to a transient stimulus. These studies have been mostly performed by the
microgravity community, the goal being to determine the gravity detection
threshold \cite{volkmann1998,laurinavicius1998,driss2008}. The procedure is the following.
A plant is exposed to an effective gravity $g_{\rm eff}$ perpendicular to its axis during a time lapse $t_0$, and then put back in
a zero gravity condition, either in space or in a clinostat. The plant then
actively bends up in response to this transient stimulus. The amplitude of the
response appears to increase when increasing the gravity intensity $g_{\rm eff}$
or when increasing the time of exposure $t_0$. Data suggest that the
response is proportional to the dose, i.e.\ to the product of the gravity intensity by the
time of exposure $g_{\rm eff} \times t_0$, a result known as the ``reciprocity rule".
Because the goal was the determination of the minimal dose to get a response,
most of the experiments have been performed at a low level of gravity ($g_{\rm eff}< g_{earth}$).

\begin{figure*}
	\centering
	\includegraphics[width=0.650\textwidth]{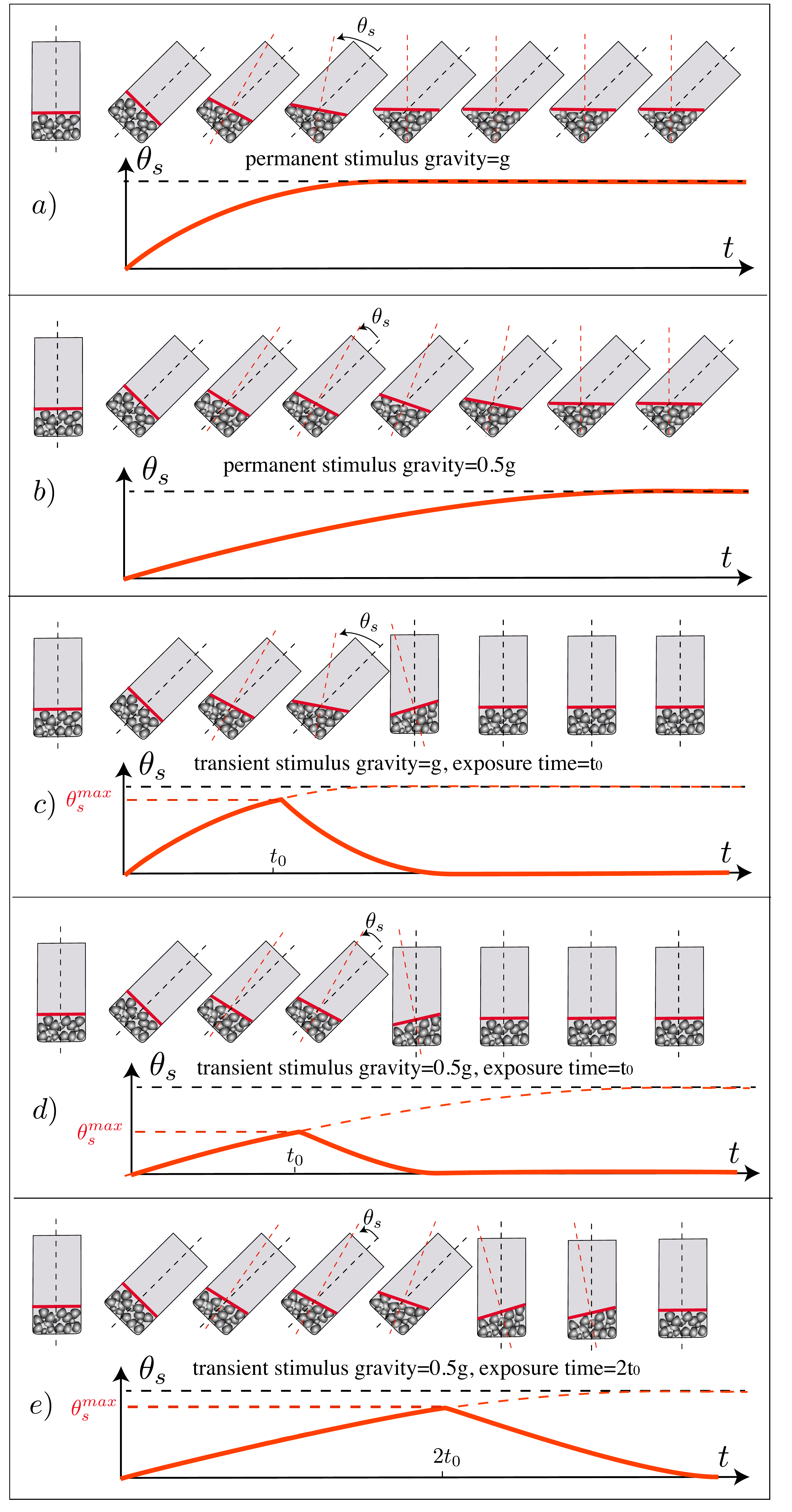}
	\caption{ Sketch of the statoliths avalanche dynamics in different
		experiments. The angle $\theta_s$ made by the free surface of the statoliths
		pile with the base of the cell is plotted as a function of time; (a) and (b)
		permanent inclination at two different gravity intensities $g$. (c), (d) and (e)
		transient inclination at two different gravity intensities $g$ and two
		exposition times}
	\label{fig:gt}
\end{figure*}

The ``reciprocity rule" explicitly involves the gravity intensity and appears a
priori  incompatible with the ``sine law" presented in section 2 showing that
the response to a permanent stimulus is independent of $g_{\rm eff}$. This incompatibility  may seriously question the position sensor hypothesis.
However, the sine law and the reciprocity rule originate from two very different types
of experiments. The first one corresponds to a permanent stimulus (gravity is
continuously present and the base of the plant remains inclined during the whole
experiment), the second one corresponds to transient stimulus during a finite
time. Here we show that the two types of procedure may be reconciled when
thinking in terms of statolith avalanches as illustrated in Fig.~\ref{fig:gt}.
When a plant is inclined, the final and steady position of the statoliths in the
lower corner of the cell is independent of the gravity intensity, providing an
explanation for the sine law as discussed above. Only the time necessary for
the statoliths to flow and to reach their final position varies (linearly) with
$g_{\rm eff}$ as sketched in Fig.~\ref{fig:gt}a and \ref{fig:gt}b. 
As long as this avalanche time in achieving the final position is negligible compared to the
time lag for the onset of the next steps in the signal transduction pathway, it has negligible effect on the observed gravitropic response. 

During a transient stimulus, when the plant is inclined during a  time $t_0$, the statoliths may not reach their final position if $t_0$ is shorter than the
time of avalanche. In this transient regime, the maximum excursion of the grains
is expected to be proportional to their velocity $V$ times the time $t_0$. The
velocity $V$ results from the balance between the gravity $M_b g_{eff}$  and the
viscous drag  $\eta V D$ ($\eta$ being the viscosity of the cytosol and $D$ the
typical size of the pile) and is thus proportional to the gravity $V=M_b g_{\rm eff} / \eta D$.  As a result, the maximum excursion should scale as
$M_b g_{\rm eff} t_0 / \eta D$, and the maximum deflection angle of the free
surface would be proportional to $g_{\rm eff} \times t_0$. This is illustrated in Fig.~\ref{fig:gt}c, d and e. When comparing two plants inclined during the same time
$t_0$ but under two different gravity $g_{eff}=g$ (Fig.~\ref{fig:gt}c) and
$g_{eff}=0.5g$ (Fig.~\ref{fig:gt}d), the excursion of the statoliths  is two times shorter for the lower gravity (the excursion being measured by $\theta_s^{max}$,
i.e.\ the deflection of the free surface of the pile compared to the lower side of
the cell). However, if the time of excursion is also multiplied by two (Fig.~\ref{fig:gt}e) to get the same dose $g_{eff}\times t_0$, one recovers approximately
the same excursion $\theta_s^{max}$. If the position is the relevant stimulus as
stipulated by the position sensor hypothesis one thus recovers the reciprocity
rule: the response is proportional to $\theta_s^{max}$ and thus to $g_{\rm eff}\:
t_0$. In experiments where plants are put in zero gravity after the exposure
time instead of coming back to the vertical, one can imagine that the statoliths will leave the membrane when $g_{\rm eff}$ is zero (as shown in lentil root statocytes \cite{driss2000}), leading to the same phenomenology. 

This interpretation of the ``reciprocity rule'' based on the avalanche dynamics only holds if the time for the detection of the statolith positions and for the signaling is sufficiently short compared to the avalanche time.  Indeed, the response at the plant scale is controlled by the slower process at the microscopic scale. Thus, the response to transient stimuli reflects the statolith avalanche dynamics only if detection and signaling processes are faster.
This is expected to be true in the low gravity conditions corresponding to the microgravity experiments exhibiting the reciprocity rule, but it will perhaps no longer be true for Earth gravity level, at least when dense statoliths are involved. 

In the ``Position Sensor Hypothesis'', the reciprocity rule would then be the
signature of the physical dynamics of  statolith displacement and not an
intrinsic gravisensitive response of the plant. These considerations
show that there is a need to perform detailed experiments under
transient stimuli, looking both at the response at the plant scale under various
levels of gravity intensity, and at the cell scale investigating the avalanche
dynamics of the statoliths.  

\subsection{The starch-less and starch-excess mutants}

The last facts usually presented in favour of the ``gravity-force sensing hypothesis" are  the effect of changing the overall mass of the amyloplasts acting as statoliths in the statocytes.  This can be achieved either through changing the growth conditions (e.g.\ light vs dark) or through the use of starch-depleted or starch-excess mutants. Kiss and coworkers used these two methods in a series of quantitative studies of the cellular and macroscopic responses (reviewed in \cite{vitha2007}).  They considered five genotypes of  \textit{Arabidopsis thaliana}: the wild type (WT), a starch-excess mutant (\textit{sex1}), two reduced starch mutants (ACG 20 and ACG 27), and a starch-less mutant (ACG 21).  From their observations, they concluded that a correlation exists between the mass or volume of the statoliths and the gravitropic response. We have re-analyzed their data to check that their conclusion still holds with the dimensionless gravitropic response $\tilde \Delta$ (equation \ref{eqn:graviresponseadim}) introduced in section 2, since mutation or treatment conditions may affect the growth rate. We have also extracted an estimate of the volume of the statoliths by measuring their apparent area in the published pictures.  Our analysis has been limited to the dark-grown (dg) hypocotyls of the starch-deficient mutants \cite{kiss1997} and to the light-grown (lg) hypocotyls of the strach-excess mutants \cite{vitha2007} for which sufficient data were published. Data are reported in table~\ref{tab:deltamutant}. We find that the plant response $\tilde \Delta$ seems to increase with the volume of the amyloplasts inside statocytes. 

\begin{table*}
	\caption{ Quantitative estimates of the amyloplasts and macroscopic response of the starch mutants in arabidopsis hypocothyl. For dark-grown material, the sizes of amyloplasts  (in terms of apparent area) range from $6.5$ (for WT$_{\rm dg}$) to $1.5\, \mu\rm m^2$ (for ACG27$_{\rm dg}$) and the starch content of each amyloplast compared to the WT$_{\rm dg}$ was slightly reduced in the ACG20$_{\rm dg}$, reduced in the ACG27$_{\rm dg}$, and the ACG21$_{\rm dg}$ amyloplasts were completely deprived of  starch. For light-grown material, the size  was larger than the dark-grown ones: the WT$_{\rm lg}$ amyloplasts are about two times bigger than the  WT$_{\rm dg}$ and the \textit{sex1}$_{\rm lg}$ are two times bigger than the WT$_{\rm lg}$. But the light also largely reduces the amount of amyloplasts for the WT$_{\rm lg}$. Taken together these genotypes can be ranked on the volume and mass of their amyloplasts as follows: WT$_{\rm dg} >$ACG20$_{\rm dg} >$ACG27$_{\rm dg} \gg$ ACG21$_{\rm dg}$ and  WT$_{ dg} \sim$ \textit{sex1}$_{\rm lg} \gg$ WT$_{\rm lg}$.  The plant response $\tilde \Delta$ seems to increase with the volume of the amyloplasts inside statocytes. The $\tilde \Delta$ of the light-grown WT$_{\rm lg}$ was 6 times lower than the one of dark-grown WT$_{\rm dg}$, and the $\tilde \Delta$ of the light-grown \textit{sex1}$_{\rm lg}$ was more than 4 times higher than WT$_{\rm lg}$ restoring the level of the $\tilde \Delta$ found for the dark-grown WT$_{\rm dg}$. It may also be noted that the value for $\tilde \Delta$  for the WT$_{\rm dg}$ is similar to the one reported for inflorescences of the same genotype by \cite{chauvet2016}. 
	}
	
	\begin{tabular}{@{}p{1.5cm}p{1.5cm}p{2.cm}p{1.5cm}p{1cm}p{1cm}p{1.1cm}p{1cm}p{0.5cm}p{1.2cm}p{0.5cm}}
		\br 
		genotype & growth conditions & Plastid size ($\mu \rm{m}^2$)\newline and starch content & Relative plastid size& $\theta_{init}$ (rad) & $\rm{d} \theta/\rm{d} t^{**}$ (rad/h) & $\rm{dL}/\rm{d} t$ (mm/h) & $R^*$  (mm)& $\tilde{\Delta}$ &$\tilde{\Delta} / \tilde{\Delta}_{\rm WT}$ & ref \\
		\mr 
		WT (WS)& Dark-grown&$6.5^*$,\newline normal&1& $\pi/2$ &0.57&0.33  & 0.4 &0.68 &1& \cite{kiss1997} \\
		
		ACG20 (WS)&Dark-grown& $4.5^*$,\newline slightly-reduced & 0.70 & $\pi/2$ & 0.42 & 0.28 & 0.4 & 0.60 & 0.88 &\cite{kiss1997}\\
		
		ACG27 (WS)&Dark-grown& $1.5^*$,\newline  reduced and  smaller&0.23 &$\pi/2$ & $0.37$ &$0.28$ &$0.4$ &$0.53$ &$0.78$&\cite{kiss1997}\\
		
		ACG21 (WS)&Dark-grown&$1.5^*$,\newline no  starch &0.24 &$\pi/2$& 0.03&0.25  &0.4  &0.05  & 0.08 &\cite{kiss1997} \\
		\mr
		WT (Col0) &Light-grown&$14.4\pm 0.8$,\newline  very few &1 &$\pi/2$ &  0.03&0.05  &0.22  &0.12  & 1 &\cite{vitha2007}   \\
		
		\textit{sex1} (Col0) &Light-grown &$30.9\pm 0.9$  &2.14  &$\pi/2$& 0.06 &0.03  &0.28  &0.52 &4.4   &\cite{vitha2007} \\ 
		\br
	\end{tabular} 

* measured from plates using Fiji, **  measured from the kinetics graphs.

	\label{tab:deltamutant}
\end{table*}

\begin{figure}
	\centering
	\includegraphics[width=0.80\linewidth]{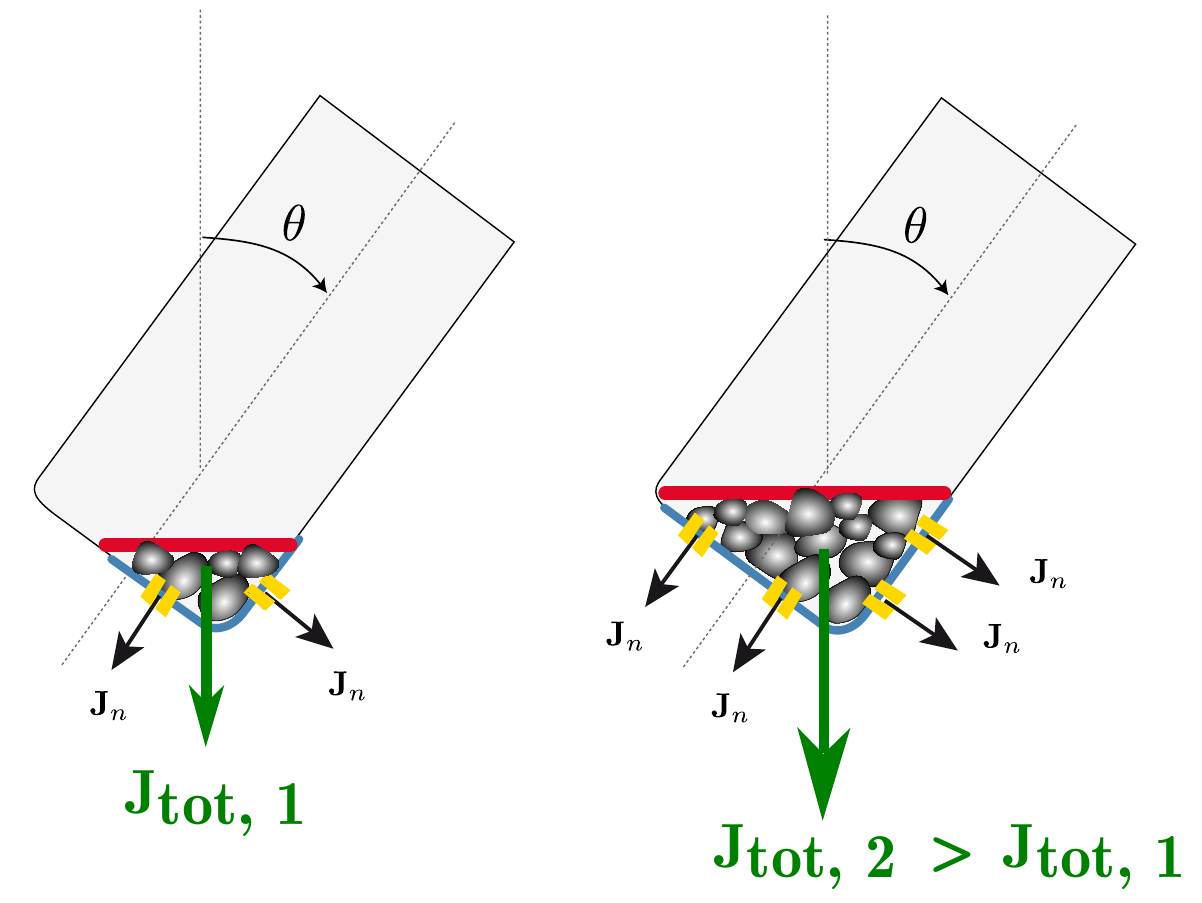}
	\caption{Sketch illustrating the effect of the statoliths volume variation on the resultant total flux of auxin for the ``position sensor hypothesis''. }
	\label{fig:starchmut}
\end{figure}

One difficulty in interpreting these experiments is that changing the growth conditions or the genotype not only affects the size of the statoliths but also their density and number. Nevertheless, the global trend supports the idea that varying the overall mass and weight of the starch-statolith changes the gravitropic sensing, in agreement with the ``gravity-force model". However, these results are not a priori incompatible with the  ``position-sensor hypothesis" we propose. Fig.~\ref{fig:starchmut} shows that increasing the amount of statoliths at a given inclination increases the contact area between the pile and the lateral side of the cell. If one assumes that the relevant signal triggering the gravitropic response is the asymmetry of the contact area as proposed in Fig.~\ref{fig:auxin}, one may expect an influence of the volume of the statolith pile on the response. 
The ``position-sensor hypothesis" could then be compatible with the macroscopic responses that have been observed by Kiss and coworkers. At this stage, it is nevertheless difficult to discriminate between the different scenarios. Further studies on mutants analysing both the plant kinematic response and the statolith properties and dynamics would provide precious information.

\section{Conclusion}  
Gravitropism, more than a century after Darwin, remains an active subject. In this perspective paper we have discussed the results of recent experiments on
the response of plant shoots to inclination at different gravity intensities.
The observation that the response is independent of the gravity intensity
strongly suggests that the gravity sensor in plants behaves like a clinometer rather than an accelerometer. We propose a new scenario in which the gravity sensors,
the statocytes, behave like position sensors, being sensitive to the position of
the statoliths within the cell. 

In this ``position sensor hypothesis" the motion of
the  statoliths is facilitated by the activity of the cytoskeleton, which may
explain the high sensitivity of plants even at small inclination angle.  This
scenario also provides a coherent framework to interpret and reconcile the
response to permanent stimuli (``the sine law"), as well as the response to
transient stimuli (``the reciprocity rule"), or to mutations in the starch content of statoliths. 
The ``position sensor hypothesis" has also important implications on the different molecular hypotheses of gravitropic perception. 
On one hand, it discards mechanisms based on the measure of forces, such as the detection of the statolith weight by the actin network or the endoplasmic reticulum, and the "protoplast pressure hypothesis". On the other hand it is compatible with the mechanisms involving the proximity of the statoliths to subcellular elements (endoplasmic reticulum \cite{strohm2014}, actin cytoskeleton \cite{toyota2013}) or a change of the intracellular trafficking due to statolith asymmetric distribution within the cell. 

 At this stage, all the conjectures
discussed in the paper remain to be confirmed or contradicted, which should
motivate more detailed analysis coupling experiments at different scales:  at
the plant scale to analyze the kinematics, at the cell scale to progress in our
understanding of  the motion of the statoliths, and at the molecular scale to
disentangle the signaling pathway. 

A last crucial remark is that a better understanding of the gravisensing is not
sufficient to apprehend and describe  the gravitropic movement and more
generally the  control of the posture of plants. Proprioception, i.e.\ the
ability of plants to feel their own curvature  leading to the tendency of plants
to unbend independently of gravitational stimulus, also plays an important role. The
development of a recent model combining  gravisensing and proprioception to
predict  the gravitropic macroscopic motion opens new perspectives and could
serve as a base to develop more elaborate models taken into account the details
of the gravisensing chain discussed in this paper \cite{hamant2016}.

\section*{Acknowledgements}

This work was supported by the CNRS PIR Interface program, ANR Blanc Grap2
(ANR-13-BSV5-0005-01), Labex MEC (ANR-10-LABX-0092), the A*MIDEX project
(ANR-11-IDEX-0001-02) funded by the French government program
\textit{Investissements d'avenir}, the European Research Council under the European Union Horizon 2020
Research and Innovation programme (grant agreement No 647384) and the French Space Agence (CNES) for F.B and
V.L.

\section*{References }

\begin{thebibliography}{10}

\bibitem{bastien2013}
Bastien R, Bohr T, Moulia B and Douady S 2013 {\em Proc. Natl. Acad. Sci.
	U.S.A.\/} {\bf 110} 755--760

\bibitem{firn1980}
Firn R~D and Digby J 1980 {\em Annu. Rev. Plant Physiol.\/} {\bf 31} 131--148

\bibitem{whippo2009}
Whippo C~W and Hangarter R~P 2009 {\em Am. J. Bot.\/} {\bf 96} 2115--2127

\bibitem{moulia2009}
Moulia B and Fournier M 2009 {\em J. Exp. Bot.\/} {\bf 60} 461--486

\bibitem{morita2010}
Morita M~T 2010 {\em Annu. Rev. Plant Biol.\/} {\bf 61} 705--720

\bibitem{gilroy2008}
Gilroy S 2008 {\em Curr. Biol.\/} {\bf 18} R275--R277

\bibitem{perbal1976}
Perbal G and Perbal P 1976 {\em Physiol. plant.\/} {\bf 37} 42--48

\bibitem{perbal2002}
Perbal G, Jeune B, Lefranc A, Carnero-Diaz E and Driss-Ecole D 2002 {\em
	Physiol. Plant.\/} {\bf 114} 336--342

\bibitem{fukaki1998}
Fukaki H, Wysocka-Diller J, Kato T, Fujisawa H, Benfey P~N and Tasaka M 1998
{\em The Plant J.\/} {\bf 14} 425--430

\bibitem{blancaflor1998}
Blancaflor E~B, Fasano J~M and Gilroy S 1998 {\em Am. J. Plant Physiol.\/} {\bf
	116} 213--222
	
\bibitem{gerttula2015}
Gerttula   S, Zinkgraf,  M, Muday G, Lewis D, and Ibatullin F~M, Brumer H,  Hart F, Mansfiel S~D, Filkov V and Groover A 2015 {\em The Plant Cell\/} {\bf 27} 2800--2813


\bibitem{vanneste2009}
Vanneste S and Friml J 2009 {\em Cell\/} {\bf 136} 1005--1016

\bibitem{nemec1900}
N{\`e}mec B 1900 {\em Deutsch. Bot. Ges., XVIII, S\/} {\bf 241--245}

\bibitem{haberlandt1900}
Haberlandt G 1900 {\em Ber. Dtsch. Bot. Ges.\/} {\bf 18} 26l--272

\bibitem{pickard1966}
Pickard B~G and Thimann K~V 1966 {\em The J. Gen. Physiol.\/} {\bf 49}
1065--1086

\bibitem{kiss1989}
Kiss J~Z, Hertel R and Sack F~D 1989 {\em Planta\/} {\bf 177} 198--206

\bibitem{weise1999}
Weise S~E and Kiss J~Z 1999 {\em Int. J. Plant Sci.\/} {\bf 160} 521--527

\bibitem{fitzelle2001}
Fitzelle K~J and Kiss J~Z 2001 {\em J. Exp. Bot.\/} {\bf 52} 265--275

\bibitem{kuznetsov1997}
Kuznetsov O~A and Hasenstein K~H 1997 {\em J. Exp. Bot.\/} {\bf 48} 1951--1957

\bibitem{hasenstein1999}
Hasenstein K~H and Kuznetsov O~A 1999 {\em Planta\/} {\bf 208} 59--65

\bibitem{toyota2013}
Toyota M, Ikeda N, Sawai-Toyota S, Kato T, Gilroy S, Tasaka M and Morita M~T
2013 {\em The Plant J.\/} {\bf 76} 648--660

\bibitem{shen1970}
Shen-Miller J 1970 {\em Planta\/} {\bf 92} 152--163

\bibitem{barlow1995}
Barlow P~W 1995 {\em Pant Cell Environ.\/} {\bf 18} 951--962

\bibitem{darwin1903}
Darwin F 1903 {\em Nature\/} {\bf 67} 571--572

\bibitem{leitz2009}
Leitz G, Kang B~H, Schoenwaelder M~E and Staehelin L~A 2009 {\em The Plant
	Cell\/} {\bf 21} 843--860

\bibitem{hasenstein2009}
Hasenstein K~H 2009 {\em Gravitational and Space Res.\/} {\bf 22} 21–33

\bibitem{yoder2001}
Yoder T~L, Zheng H~q, Todd P and Staehelin L~A 2001 {\em Am. J. Plant
	Physiol.\/} {\bf 125} 1045--1060

\bibitem{perbal2003}
Perbal G and Driss-Ecole D 2003 {\em Trends Plant Sci.\/} {\bf 8} 498--504

\bibitem{strohm2014}
Strohm A~K, Barrett-Wilt G~A and Masson P~H 2014 {\em Front. Plant Sci.\/} {\bf 5} 148

\bibitem{kiss1997}
John~Z K, Mary~M G, Allison~J M and Kathi~S S 1997 {\em Plant Cell Physiol.\/}
{\bf 38} 518--525

\bibitem{fasano2001}
Fasano J~M 2001 {\em The Plant Cell\/} {\bf 13} 907--922

\bibitem{joo2001}
Joo J~H, Bae Y~S and Lee J~S 2001 {\em Am. J. Plant Physiol.\/} {\bf 126}
1055--1060

\bibitem{toyota2008}
Toyota M, Furuichi T, Tatsumi H and Sokabe M 2008 {\em Am. J. Plant Physiol.\/}
{\bf 146} 505--514

\bibitem{kolesnikov2015}
Kolesnikov Y~S, Kretynin S~V, Volotovsky I~D, Kordyum E~L, Ruelland E and
Kravets V~S 2015 {\em Protoplasma\/} {\bf 253} 987--1004

\bibitem{band2012}
Band L~R, Wells D~M, Larrieu A, Sun J, Middleton A~M, French A~P, Brunoud G,
Sato E~M, Wilson M~H, Peret B, Oliva M, Swarup R, Sairanen I, Parry G, Ljung
K, Beeckman T, Garibaldi J~M, Estelle M, Owen M~R, Vissenberg K, Hodgman T~C,
Pridmore T~P, King J~R, Vernoux T and Bennett M~J 2012 {\em Proc. Natl. Acad.
	Sci. U.S.A.\/} {\bf 109} 4668--4673

\bibitem{petravsek2006}
Petr{\'a}{\v s}ek J, Mravec J, Bouchard R, Blakeslee J~J, Abas M,
Seifertov{\'a} D, Wi{\'s}niewska J, Tadele Z, Kube{\v s} M, {\v C}ovanov{\'a}
M, Dhonukshe P, Sk{\r u}pa P, Benkov{\'a} E, Perry L, K{\v r}e{\v c}ek P, Lee
O~R, Fink G~R, Geisler M, Murphy A~S, Luschnig C, Za{\v z}{\'\i}malov{\'a} E
and Friml J 2006 {\em Science\/} {\bf 312} 914--918

\bibitem{wisniewska2006}
Wi{\'s}niewska J, Xu J, Seifertov{\'a} D, Brewer P~B, Ru{\v{z}}i{\v{c}}ka K,
Blilou I, Rouqui{\'e} D, Benkov{\'a} E, Scheres B and Friml J 2006 {\em
	Science\/} {\bf 312} 883--883

\bibitem{friml2002}
Friml J, Wi{\'s}niewska J, Benkov{\'a} E, Mendgen K and Palme K 2002 {\em
	Nature\/} {\bf 415} 806--809

\bibitem{kleine2010}
Kleine-Vehn J, Ding Z, Jones A~R, Tasaka M, Morita M~T and Friml J 2010 {\em
	Proc. Natl. Acad. Sci. U.S.A.\/} {\bf 107} 22344--22349

\bibitem{rakusova2011}
Rakusov{\'{a}} H, Gallego-Bartolom{\'{e}} J, Vanstraelen M, Robert H~S,
Alabad{\'{\i}} D, Bl{\'{a}}zquez M~A, Benkov{\'{a}} E and Friml J 2011 {\em
	The Plant J.\/} {\bf 67} 817--826

\bibitem{dumais2013}
Dumais J 2013 {\em Proc. Natl. Acad. Sci. U.S.A.\/} {\bf 110} 391--392

\bibitem{iino1996}
Iino M, Tarui Y and Uematsu C 1996 {\em Plant, Cell and Environment\/} {\bf 19}
1160--1168

\bibitem{galland2002}
Galland P 2002 {\em Planta\/} {\bf 215} 779--784

\bibitem{brown1995}
Brown A~H, Chapman D~K, Johnsson A and Heathcote D 1995 {\em Physiol. Plant.\/}
{\bf 95} 27--33

\bibitem{chauvet2016}
Chauvet H, Pouliquen O, Forterre Y, Legu\'{e} V and Moulia B 2016 {\em Sci.
	Rep.\/} {\bf 6} 35431

\bibitem{tanimoto2008}
Tanimoto M, Tremblay R and Colasanti J 2008 {\em Plant Mol. Biol.\/} {\bf 67}
57--69

\bibitem{myers1995}
Myers A~B, Glyn G~H, Digby J and Firn R~D 1995 {\em Ann. Bot.\/} {\bf 75}
277--280

\bibitem{hamant2016}
Hamant O and Moulia B 2016 {\em New Phytol.\/} {\bf 212} 333--337

\bibitem{bastien2014}
Bastien R, Douady S and Moulia B 2014 {\em Front. Plant Sci.\/} {\bf 5} 136 

\bibitem{galston1949}
Galston A~W and Hand M~E 1949  {\em Am. J. Bot.} {\bf 36} 85--94

\bibitem{hopkins2008}
Hopkins W~G and H\"uner N~P~A 2009 {\em Introduction to plant physiology, 4th edition} (Wiley)

\bibitem{silk1984}
Silk W~K 1984 {\em Ann. Rev. Plant Physiol.\/} {\bf 35} 479--518

\bibitem{bastien2016}
Bastien R, Legland D, Martin M, Fregosi L, Peaucelle A, Douady S, Moulia B and
H{\"o}fte H 2016 {\em The Plant J.\/} {\bf 88} 468--475

\bibitem{andreotti2013}
Andreotti B, Forterre Y and Pouliquen O 2013 {\em Granular media: between fluid
	and solid\/} (Cambridge University Press)

\bibitem{sack1985}
Sack F~D, Suyemoto M~M and Leopold A~C 1985 {\em Planta.\/}  {\bf 165} 295--300

\bibitem{sack1986}
Sack F~D, Suyemoto M~M and Leopold A~C 1986 {\em Am. J. Bot.\/}  {\bf 73} 1692--1698

\bibitem{saito2005}
Saito C, Morita M~T, Kato T and Tasaka M 2005 {\em The Plant Cell\/} {\bf 17}
548--558

\bibitem{nakamura2011}
Nakamura M, Toyota M, Tasaka M and Morita M~T 2011 {\em The Plant Cell\/} {\bf
	23} 1830--1848

\bibitem{zheng2015}
Zheng Z, Zou J, Li H, Xue S, Wang Y and Le J 2015 {\em Mol. plant\/} {\bf 8} 660--663

\bibitem{guo2014}
Guo M, Ehrlicher A~J, Jensen M~H, Renz M, Moore J~R, Goldman R~D,
Lippincott-Schwartz J, Mackintosh F~C and Weitz D~A 2014 {\em Cell\/} {\bf
	158} 822--832

\bibitem{Blancaflor2013}
Blancaflor E~B 2013 {\em Am. J. Bot.\/} {\bf 100} 143--152

\bibitem{yamamoto2002}
Yamamoto K and Kiss J~Z 2002 {\em Am. J. Plant Physiol.\/} {\bf 128} 669--681

\bibitem{hou2004}
Hou G, Kramer V~L, Wang Y~S, Chen R, Perbal G, Gilroy S and Blancaflor E~B 2004
{\em The Plant J.\/} {\bf 39} 113--125

\bibitem{mancuso2006}
Mancuso S, Barlow P~W, Volkmann D and Baluška F 2006 {\em Plant Signaling
	Behav.\/} {\bf 1} 52--58

\bibitem{friedman2003}
Friedman H, Vos J~W, Hepler P~K, Meir S, Halevy A~H and Philosoph-Hadas S 2003
{\em Planta\/} {\bf 216} 1034--1042

\bibitem{volkmann1999}
Volkmann D, Balu{\v{s}}ka F, Lichtscheidl I, Driss-Ecole D and Perbal G 1999
{\em The FASEB J.\/} {\bf 13} 143--147

\bibitem{dharmasiri2006}
Dharmasiri S, Swarup R, Mockaitis K, Dharmasiri N, Singh S, Kowalchyk M,
Marchant A, Mills S, Sandberg G, Bennett M {\em et~al.\/} 2006 {\em
	Science\/} {\bf 312} 1218--1220

\bibitem{volkmann1998}
Volkmann D and Tewinkel M 1998 {\em Adv. Space Res.\/} {\bf 21} 1209--1217

\bibitem{laurinavicius1998}
Laurinavicius R, Svegzdiene D, Buchen B and Sievers A 1998 {\em Adv. Space
	Res.\/} {\bf 21} 1203--1207

\bibitem{driss2008}
Driss-Ecole D, Legu{\'{e}} V, Carnero-Diaz E and Perbal G 2008 {\em Physiol.
	Plant.\/} {\bf 134} 191--201

\bibitem{driss2000}
Driss-Ecole D, Jeune B, Prouteau M, Julianus P and Perbal G 2000 {\em Planta\/}
{\bf 211} 396--405

\bibitem{vitha2007}
Vitha S, Yang M, Sack F~D and Kiss J~Z 2007 {\em Am. J. Bot.\/} {\bf 94}
590--598
\end{thebibliography}


\end{document}